\documentstyle[12pt]{article}
\input{tcilatex}

\begin{document}

\title{Galactic Gamma halo by heavy neutrino annihilations ?}
\author{D. Fargion \thanks{%
Phys. Dept. University of Rome "La Sapienza", Italy; \newline
E.E.Faculty Technion Institute, Haifa, Israel}, \and R. Konoplich \thanks{%
Center for Cosmoparticle Physics - "Cosmion", Moscow, Russia. \newline
Institute of Applied Mathematics "M. V. Keldysh", Moscow, Russia. \newline
Moscow Engineering Physics Institute.}, \and M. Grossi \thanks{%
Phys. Dept. University of Rome "La Sapienza", Italy.}, \and M. Khlopov $%
^{\dag}$.}

\maketitle

\begin{abstract}
The diffused gamma halo around our Galaxy recently discovered by EGRET could
be produced by annihilations of heavy relic neutrinos N (of fourth
generation), whose mass is within a narrow range ($M_Z/2 < m_N < M_Z$).
Neutrino annihilation in the halo may lead to \newline
either ultrarelativistic electron pairs whose Inverse Compton Scattering on
infrared and optical galactic photons could be the source of observed GeV
gamma rays, or prompt 100 MeV - 1 GeV photons (due to neutral pion
secondaries) born by $N \bar{N} \rightarrow Z \rightarrow q \bar{q}$
reactions.\newline
The consequent gamma flux ($10^{-7} \,-\, 10^{-6} \, cm^{-2}\,s^{-1} \,
sr^{-1}$) is well comparable to the EGRET observed one, and it is also
compatible with the narrow window of neutrino mass $45\, GeV < m_N < 50\,
GeV $, recently required to explain the underground DAMA signals.\newline
The presence of heavy neutrinos of fourth generation do not contribute much
to solve the dark matter problem of the Universe, but may be easily
detectable by outcoming LEP II data.
\end{abstract}

\section{Introduction}

The surprising recent discover of a diffused galactic gamma halo
[1] could be explained by the presence at high galactic latitudes
either of a diffused high energy cosmic ray sources (for instance
fast running pulsars "Geminga" like [2]), or by the presence of
neutral molecular clouds [3,4] at large galactic latitude, where
high energy protons (tens of GeV) collide.\newline However these
two (ad hoc) solutions are not widely accepted and are not the
unique ones.\newline The presence of heavy relic neutrinos of new
fourth generation in the galactic halo and their annihilations may
offer an elegant solution to the gamma ray halo origin.\newline

Actually a fashionable candidate of a cold dark matter scenario is the
neutralino, the lightest stable supersymmetric particle in the Minimal
Supersymmetric Standard Model. However it doesn't seem that neutralino could
be in general an acceptable candidate as a source of gamma rays in the halo
[1], mostly because its annihilations in light fermion pairs, due to its
Majorana nature, are required to be in the $p$ wave, as a consequence of
Fermi statistics. At low energy the cross section is suppressed by a factor $%
\beta^2$ ($\beta$ is neutralino velocity in the Galaxy), in comparison with $%
s$ wave, and the resulting gamma flux could not be comparable with
observations (below several order of magnitude) unless we
introduce a model of a clumpy galactic halo, whose clustering
processes is arbitrary and mysterious.\newline However being
$\sigma \sim m^2_{\chi} [\beta^2 + (m_f/m_{\chi})^2]$,
annihilations in heavy fermions ($c\bar{c},\,b\bar{b},\,t\bar{t}$
and the channels $W^+W^-,\,ZZ,\,gg$) are not $\beta^2$ suppressed,
but the resulting gamma flux (obtained by Monte Carlo simulations)
would still be less than $ 10^{-8}\,cm^{-2}\,s^{-1}\,sr^{-1}$ in
the range of masses with $0.025 < \Omega_{\chi} < 0.5\,$
[5].\newline On the contrary heavy Dirac neutrino of a fourth
generation, at masses $m_N > M_Z/2$, while not being able to solve
all the dark matter problem in the galactic halo, could annihilate
(more easily than neutralino) into fermions by $s$ wave channel
(with no $\beta^2$ suppression). Their relics produce either
directly gammas by neutral secondary pions, or gamma rays by ICS
of relativistic electrons (primaries or as secondary decay
products of heavier leptons in the annihilation chains $N\bar{N}
\rightarrow l\bar{l} \rightarrow e^+ e^-$) onto thermal photons
($IR$, $optical$) near and above the galactic plane.\newline The
estimated flux we derived here is roughly close to EGRET results
in the hypothesis of a smooth and homogeneous galactic
halo.\newline Moreover heavy neutrino annihilations may also
produce long life antiprotons (as well as unstable antineutrons),
whose further annihilations on common protons (gas, plasma,
molecular clouds), may also produce hundreds MeV neutral pions and
secondary gamma photons. Anyway this additional possibility would
require both neutrino solution and baryons in the halo, and it
might be less plausible.\newline Our heavy neutrino candidate is
not an ad hoc model; its unique free parameter, the neutrino mass,
is already constrained by cosmological data (no high pollution of
$e^+ e^-$ cosmic rays in the range $M_Z < m_N < 300 GeV $ is
observed), while recent early data from underground detector of
WIMP could be attributed to heavy neutrinos, streaming in DAMA
detector at a very narrow and defined mass window $45 \, GeV < m_N
< 50 GeV$ [6].

\section{Heavy neutrino cosmological evolution}

The early Universe provide a unique laboratory to test elementary
particles which are produced in the cosmic thermal bath at huge
rate and at highest energies. The number of known lepton families
known is actually fixed at three. Their cosmic relics, namely the
tiny survived fraction of baryons (over antibaryons) and the
corresponding tiny fraction of leptons (electrons) over
antileptons are a minor trace with respect to the huge number
density of massless gauge boson relics, the 2.75 BBR photons,
whose number density is comparable with the expected electron,
muon and tau neutrinos. These ''light'' neutrinos in thermal
equilibrium at very early stages of Universe evolution, may share
a small mass, as last experimental evidences from Superkamiokande
(atmospheric neutrino) recently suggest, making ''light''
neutrinos a preferable candidate for hot dark matter in the
Universe.\newline Recently it has been suggested that light $\nu $
presence in galactic halo might be already probed by the survival
of ultrahigh cosmic rays originated at high cosmic distances
(above GZK cut off), whose only explanation may be indebted to UHE
neutrino - relic galactic neutrino interactions [7, 8].\newline
Cosmological nucleosynthesis arguments bounds the total flavour
number of light neutrinos to three or four, while terrestrial data
on the Z width at
LEP infer severe constrains on any further ''light'' ($m_{\nu }\ll M_{Z}/2$%
) weakly interacting neutrino families detectable from boson gauge
decay. \newline Anyway, a fourth lepton family above $M_{Z}/2$
masses is not experimentally forbidden. A similar ''heavy'' stable
neutrino has been proposed nearly 20 years ago as a Cold Dark
Matter candidate, able to reach the critical mass density at GeV
masses (the Lee - Weinberg - Dolgov suggestion now in disagreement
with LEP limits on Z width [9]) and at TeV masses [10]
\footnote{An apparently simple electroweak model with a fourth
heavy neutrino is not \ so easy to define. First of all it has to
be included an additional quark pair in order to cancel anomalies,
and assured the stability of the lightest particle of the family
(heavy N). In a SU(5) model the dimension five operators appearing
in the lagrangian density lead to the neutral fermion decay with a
lifetime shorter than the age of the Universe ($\tau \sim
M_{PL}^{2}/m_{N}^{3} \sim M_{PL}^{2}/ M_{W}^{3}\sim 10^{8}s\ll
H_{0}^{-1}\sim 3\cdot 10^{17}s).$\\ Theoretical models predicting
four fermion families which exclude these operators contribution
can be found in Ref [11-13]. In Ref. [11] for example it is shown
that a non Abelian horizontal symmetry, broken at some high scale
can guarantee the replica of a fourth fermion generation with a
heavy stable neutrino, leaving global U(1) (related to lepton
number conservation) unbroken.\\ New fermion masses are
constrained in the range [11]

\[ M_{N} \sim 50 \,GeV \, , \;M_{L} > 80\,GeV \, ,\; M_{B'}\sim
100 \,GeV \, ,\;M_{T'}\,>\, 150\, GeV. \] }.

Let us remind that the residual cosmological abundance of heavy
neutrinos in the Universe depends on the conditions of "freeze
out", when they decouple from a state of thermal equilibrium with
other particle species during the primordial stages of Universe
evolution. \newline At high temperatures ($T \gg m_N$) heavy
neutrino concentration is comparable with that of photons. When
the temperature drops below neutrino mass ($T \approx m_N$), due
to Universe expansion,  neutrino are still in thermal equilibrium
but relic concentration decreases like $n_N \propto exp(-m_N/T)$
until the decoupling. At this moment the weak interactions rate
becomes comparable with cosmic expansion, and too slow to keep
neutrinos in equilibrium with other particles. The knowledge of
these processes cross sections is essential in order to calculate
the neutrino density today.\\ According to the Standard Model,
neutrino relic abundance is given by [14]

\begin{equation}
\ n_N \simeq \frac{2 \times 10^{-18}}{g_*^{1/2} M_p m_N (\overline{\sigma
\beta})_f} \bigg[ 40 + \ln \left( \frac{g_s}{g_*^{1/2}} M_p m_N (\overline{%
\sigma \beta})_f \right) \bigg] n_{\gamma} (T) ,
\end{equation}
\

where $g^* = N_{bos} + \frac{7}{8} N_{ferm}$ is the number of
effective degrees of freedom at temperature T, $g_s$ is the number
of particle spin states $M_p$ is the proton mass,
$(\overline{\sigma \beta})_f $ is the thermally averaged
annihilation cross section at freeze out, $n_{\gamma} = 0.24 T^3$
is the cosmic photon number density.

The main events of heavy neutrinos annihilation in the Universe
are \newline

\[
N\bar{N} \rightarrow f \bar{f},
\]
if $M_Z/2 < m_N < m_W$,
where the cross section decreases as $4m^2_N/(4m^2_N - M^2_Z)^2$ for growing $m_N$, and%
\newline

\[
N \bar{N} \rightarrow W^+ W^-
\]

if $m_N > m_W$ , where the cross section grows like
$m^2_N$.\newline Because neutrino relic density $\rho_N \propto
\sigma^{-1}$ [9-10],  it grows until a maximum value near $m_N
\sim M_W$ with $\rho_{max}/\rho_c \approx 10^{-2} $, and it starts
to decrease in the mass range $m_N > m_W$, as Enquist showed in
his work [15]. The Heavy neutrino can not reach the critical value
$\Omega = 1$, at least below a few TeV energy mass, where the
Standard Model may be applied.
\newline The neutrino "fluid" decouples from thermal equilibrium (at $T_f \approx
m_N/30$) and starts to cluster, in matter dominated Universe, at
earlier times ($z \approx 10^5$) compared with  the baryons which
remain in equilibrium with photons until $z \sim 1500$. After
recombination, baryonic matter is gravitationally captured by
primordial neutrino seeds, while heavy neutrinos loose energy
moving in the non static gravitational field of ordinary matter
collapsing. This mechanism drives neutral lepton clustering in
galactic dark halos as a consequence of gravitational interaction
with baryons.
\newline The neutrino density increase[6, 16-18] in the central part of the
Galaxy may become 5 $ \div$ 7 orders of magnitude larger than its
cosmological value \footnote{We have to note that in a Cold Dark
Matter scenario dominated by neutralino ($\Omega_{\chi} \simeq
1$), in the very particular case expecting that both $\chi$ and N
share comparable mass $and$ very similar electroweak interactions,
the mean galactic overdensity of our heavy neutrino should not
exceed the value $3 \cdot 10^4 \rho_c$. In this case we can not
explain naturally the DAMA signal and the gamma flux with no
additional local clustering.
\newline However for the most general case ($\Omega_{DM} <
0.2 - 0.5$, $and/or$ $m_{\chi} < m_N $, $and/or$ $\sigma_{\chi}
> \sigma_N$) the neutrino decoupling and clustering may occur in a
separated way more efficiently than other WIMP component [18]. In
general there is no strong constraint for a neutrino density
contrast with $\Omega_N \sim  10^{-3}$. Therefore the clustering
of neutrino could not be bounded by a twin ghost WIMP candidate
whose free parameters are totally unknown ( mass, cross-section,
helicity). Moreover there are real possibilities that baryonic
dark matter (MACHOs, molecular clouds) constitutes a relevant
component of galactic dark matter. In this scenario the density
contrast $(\delta \rho / \rho)_{bar}$ in the galactic halo is much
larger than $10^5 \div 10^7$, and the consequent neutrino
clustering may easily reach or exceed the assumed one.}. For this
reason the heavy neutrinos N may reannihilate (after the earliest
epoch of Hot Universe) with antineutrinos in the halo leading to a
flux of ordinary particles beyond the galactic plane, which may be
more easily detectable.\newline Contributing a small part of the
total dark matter density and participating rarely to
reannihilation, the primordial fourth generation neutrinos can
cause significant effects by their annihilation processes.
Electrons, positrons, nucleons (antinucleons) and gamma rays are
possible final products of the annihilation chains. Relic leptons
(namely electron pairs), interacting by ICS with soft interstellar
radiation background, and gamma photons produced by $\pi^0$ decay
could be the favorite sources of the GeV radiation observed in the
galactic halo.

\section{Neutrino annihilation in relativistic electron pairs: Inverse
Compton scattering on the galactic interstellar radiation as a source of
gamma rays.}

Heavy neutrinos could directly annihilate in relativistic electron pairs ($N%
\bar{N} \rightarrow Z \rightarrow e^+ e^-$) or through secondary
decay processes of heavier particles as it is showed in Table 1,
where we considered only the annihilations leading to most
energetic electron pairs.  We indicated with $\Phi$ the electron
pair normalized probability production (through each Z decay) for
each corresponding chain channels (direct $\phi_e$, or via $\mu$
or $\tau$ decays $\phi_{\mu e}$, $\phi_{\tau e}$, $\phi_{\tau \mu
 e}$).
 Charged pions and neutrons generated in
jets by annihilation in quark - antiquark pairs through Z or W
hadronic decay also give electrons as secondaries, but their
energy is lower than one order of magnitude  .
\newline The complete reaction chains at lowest energies are not
considered here, but we use it as a first reference as well as the
detailed Monte Carlo processes [19].\newline The presence of heavy
neutrino would determine (as a consequence of occasional
annihilations) a flux of cosmic rays in the halo of the Galaxy
whose intensity should be in the range of actual detector
sensibilities. Anyway it could be hardly distinguishable from the
standard sources contribution (as supernova or supernova remnants)
to the galactic background.\newline Electrons and positrons are
trapped by magnetic fields, and propagating through the Galaxy
they loose either ''memory'' of their ''place of birth'' or energy
by bremsstrahlung, synchrotron or ICS. These processes determine a
broadening of different electron ''lines'' ($N\bar{N}\rightarrow l\bar{l}%
\rightarrow e^{+}e^{-}$), so that $e^{-}\,(e^{+})$ spectra (even considering
electrons and positrons that come from hadron decays) are at final stages
described by numerical spectra and consequent approximated power law $%
E^{-\alpha }$ [19,20]. The numerical simulations of $N\bar{N}$
annihilation performed with the package PYTHIA 5.7 [19] with
suitable modifications to include a fourth generation of fermions,
show that such $N\bar{N}$ \ relic electron fluxes are considerably
lower than observed neighbor galactic background (in the range of
masses $45\,GeV<m_{N}<M_{Z}$). Such cosmic ray  input can not be
used to confirm or refuse heavy neutrino presence in galactic
halo. Recently the possible excess of positrons has been
considered as a probe of heavy neutrino annihilations [20].
\newline A gamma signal at {\em high galactic latitudes} detected
by EGRET could be a test of neutrino annihilation in the Milky Way
halo, because no standard gamma rays sources are known in these
galactic regions ( $\gamma$-rays  can practically travel in
straight paths through the Galaxy
 with  no absorption because the mean $\gamma \,-\,p$ free path
length at typical interstellar density is about 20 Mpc
[21]).\newline In the present section we consider as source of
high energy radiation in the halo the Inverse Compton Scattering
(ICS) of relativistic electrons (by $N\bar{N}$ collisions) on
''soft'' radiative backgrounds diffused in the Galaxy, mainly
Cosmic Background Radiation (CBR), and  the Interstellar Radiation
Field (infrared and optical photons). The electron energies,
required by ICS mechanism in order to generate gamma radiation,
are given by

\begin{equation}
\ E_{\gamma} = \frac{4}{3} \epsilon_{ph} \left( \frac{E_e}{m_e c^2} \right)^2
\end{equation}
\

with $\epsilon_{ph} $ the target photon energy.\newline

The Compton scattering on cosmic microwave photons call for too
large neutrino masses ($m_N > 1 \, TeV$) in order to reach gamma
GeV energies. In this range of $m_N$ masses perturbative theory
could not be applied and neutrino
interactions could not even be weak anymore. No clear study for $N\bar{%
N}$ annihilation is known above TeV energies. Therefore we
excluded neutrinos with mass above 1 TeV from our
consideration.\newline The ICS on IR and optical photons with
respectively $E_e \geq 50 \,GeV$ or $E_e \geq 10 \, GeV$, is more
efficient in gamma rays production. Indeed assuming interstellar
radiation being represented by the following scaling law

\begin{equation}
\ n_{ph} (r) = \frac{n_{ph} (0)}{1 + r^2/a^2_{\gamma}}.
\end{equation}

where $r$ is the distance from the galactic plane, and $a_{\gamma}
= 10\,kpc$ is the characteristic length of interstellar radiation
distribution in the Galaxy, the photon density could be considered
roughly constant in a region of radius $a_{\gamma}$ [22].\\ We
have already underlined that numerical simulation of electron
pairs spectra from $N\bar{N}$ annihilation for different values of
neutrino masses ($m_N = 45,\, 50,\, 100,\,300\,GeV$) [19], could
be approximated, in a reasonable energy windows,  by a power law

\begin{equation}
\ J(E_e) = K E_e^{-\alpha} .
\end{equation}
\

where K is a normalization constant. An electron distribution with
such a power law, interacting by ICS with photons at energy
$\epsilon_{ph}$ and number density $n_{ph}$, generate radiation at
higher energy whose intensity is given by [23-25]

\begin{equation}
\ J_{\gamma} (E_{\gamma}) = \int dr \int I(E_e) dE_e \int \sigma n_{ph} (
\epsilon_{ph},\,r ) d\epsilon_{ph}
\end{equation}
\

Estimates of the power index $\alpha$ lead to the determination of
gamma intensities by ICS, leading to the following power law:

\begin{equation}
\ J_{\gamma} (E_{\gamma}) = \frac{2}{3} K a_{\gamma} n_{ph} \sigma_T \left(
\frac{\bar{\epsilon}_{ph}}{(mc^2)^2} \right)^{(\alpha -1)/2}
E_{\gamma}^{-(\alpha + 1)/2}
\end{equation}
\

where $n_f$ is the target photon background density,
$\bar{\epsilon}_{ph}$ its average energy and $\sigma_T$ is the
Thomson cross section. \newline Gamma intensity has been
calculated for $m_N = 45,\, 50,\, 100,\,300\,GeV$. The largest
fluxes have been obtained for ICS on optical photons. Assuming an
average $N\bar{N}$ clustering $\rho_N^{gal}/ \rho_N^{cosm} =
10^7$, for a neutrino mass $m_N = 50 \,GeV$, we found  a gamma
flux

\begin{equation}
\ \frac{dN_{\gamma}}{dS\, dt\, d\Omega\, dE_{\gamma}} \simeq 2 \cdot 10^{-7}
A(\psi) \left( \frac{E_{\gamma}}{GeV} \right)^{-1.55} \left( \frac{a_{\gamma}%
}{10\, kpc} \right) \, cm^{-2}\,s^{-1}\, sr^{-1}\,GeV^{-1}
\end{equation}

and for $m_N = 100\,GeV$ one finds

\begin{equation}
\ \frac{dN_{\gamma}}{dS\, dt\, d\Omega\, dE_{\gamma}} \simeq 3 \cdot 10^{-7}
A(\psi) \left( \frac{E_{\gamma}}{GeV} \right)^{-1.5} \left( \frac{a_{\gamma}%
}{10\, kpc} \right) \, cm^{-2}\,s^{-1}\, sr^{-1}\, GeV^{-1} \,.
\end{equation}

The flux depends on the angular coordinate $\psi$ which is the angle between
the line of sight $L$ and the direction of the galactic center, related to
galactic coordinates $l,\,b$ by $\cos \psi = \cos l \cos b$.\newline
$A(\psi)$ is an adimensional integral of interstellar photon density along
the corresponding line of sight, and has been defined as

\begin{equation}
\ A( \psi ) = \frac{1}{a_{\gamma}} \int_{line\, of\, sight} \frac{dr(\psi)}{%
(1 + r(\psi)^2/a^2_{\gamma})}
\end{equation}

At high galactic latitudes $A( \psi ) \geq 1$.\\
Gamma intensity due to infrared background is less abundant than
optical photons as a consequence of the spectral power law
$E^{-\alpha}$.\newline Our integral flux is comparable with EGRET
observed one, whose value at high galactic latitudes is:

\[
\Phi_{\gamma} (E > 1\, GeV) \simeq 8 \cdot 10^{-7} \, cm^{-2}\,s^{-1}\,
sr^{-1}.
\]

We found that gamma flux due to ICS  is

\[
\Phi_{\gamma} (E > 1\, GeV) \simeq 4 \cdot 10^{-7}\,A(\psi)\, \left( \frac{%
a_{\gamma}}{10\, kpc} \right) \, cm^{-2}\,s^{-1}\,sr^{-1}\, ,
\]

with $m_N \simeq 50\,GeV$, and

\[
\Phi_{\gamma} (E > 1\, GeV) \simeq 6 \cdot 10^{-7}\,A(\psi)\, \left( \frac{%
a_{\gamma}}{10\, kpc} \right) \, cm^{-2}\,s^{-1}\,sr^{-1}
\]

with $m_N \sim 100\,GeV$.\\

 It has to be underlined that the same
tens of $GeV$ electrons which collide with interstellar radiation
could scatter on the microwave background, uniformly distributed
in the Galaxy (even at high latitudes) and originate additional
radiation at peak energy $E_{\gamma} \sim 300\,keV$ with a flux
$J_{\gamma} \simeq 10^{-2} \, cm^{-2}\,s^{-1}\, sr^{-1}$ as well
as ICS of $GeV$ electrons with a flux $J_{\gamma} \sim
0.3\,cm^{-2}\,s^{-1}\, sr^{-1}$ and peak energy $E_{X} \sim
3\,keV$. Actually this radiation could not be easily observable in
this background radiation region by present detectors.\newline

A further constraint on this model could be the detection of a
radio background at high galactic latitudes due to synchrotron
losses of tens GeV electrons. Because of the same dependence on
the square of electron energy, it should be reasonable to admit
that the same electrons generating gamma radiation could also
interact with magnetic fields in the halo. Such a synchrotron
emission would be characterized by an average frequency

\begin{equation}
\ \nu = \gamma^2 \left( \frac{eB}{2 \pi m_e} \right) \sim 1\, GHz \left(
\frac{B}{1 \mu G} \right) \left( \frac{\gamma}{2 \cdot 10^4} \right)^2
\end{equation}
\

with a characteristic scale for the magnetic field B = 1 $\mu G$.%
\newline
ICS and synchrotron luminosity are related by

\begin{equation}
\frac{L_{sync}}{L_{IC}} \simeq \frac{U_{mag}}{U_{rad}}
\end{equation}

where $U_{mag}$ and $U_{rad}$ stand for the energy density of magnetic and
interstellar radiation fields, so the typical density flux of synchrotron
radiation would be

\begin{equation}
\ J_{sync} \sim 5 \cdot 10^4 \left( \frac{B}{1 \mu G} \right ) \left( \frac{%
\gamma}{2 \cdot 10^4} \right)^{-2} \left( \frac{U_{rad}}{0.2 \, eV\, cm^{-3}}
\right)^{-1} \,Jy \,.
\end{equation}
\

considering an optical background energy density (due to standard
sources in the Galaxy) comparable with microwave background in a
sphere of radius
$r = 10 \,kpc$ at a characteristic value $U_{rad} = 0.2 \,eV\,cm^{-3}$.%
\newline
It is remarkable that recent model interpretations of gamma and
radio backgrounds do indeed reach similar expectation fluxes
comparable with the observed ones [26]. In a sentence radio -
gamma halo association seems to be reliable and it confirms the
peculiar role of ICS by ultrarelativistic electrons in a wide
halo.

\section{Gamma photons from heavy neutrino annihilations}

A different route from neutrino annihilations in the halo  could also
originate a diffuse background of gamma radiation with a continuum energy
spectrum. This main source of gamma production is the decay of $\pi^0$
mesons created in the fragmentation of quarks through the annihilation
channels $N\bar{N} \rightarrow Z \rightarrow q \bar{q}$.\newline
If $m_N > m_W$ a gamma spectrum is given by W decay, because the channel $N%
\bar{N} \rightarrow W^+ W^-$ becomes dominant.

This mechanism of gamma emission depends only on neutrino
distribution in the halo, with no need of introducing neither soft
radiative backgrounds as for above ICS case, nor a molecular gas
distribution beyond galactic plane which could interact with high
energy cosmic rays (see De Paolis et al. hypothesis about MACHOs
role as gamma halo source [3,4] ).\newline In this way gamma
radiation should preserve the memory of spatial distribution
similar to dark matter in the Galaxy.\newline

The photon flux is described by the following expression:

\begin{equation}
\ J_{\gamma} = \frac{1}{4 \pi m^2_N} \sum_{i} \sigma_i v \frac{dN^i} {dE}
\int_{line\,of\,sight} \rho^2 \,(r) dr(\psi)
\end{equation}
\

where $\psi$ is the angle between the line of sight and the
galactic center, $\rho\,(r)$ is heavy neutrino density as a
function of galactocentric radius, and $\sum_i \sigma_i v
\frac{dN^i}{dE}$ counts all possible final photon channels
($\frac{dN^i}{dE}$) which could contribute to the gamma photons
emission. The integral of neutrino density along the line of sight
L depends on the halo model chosen for a dark matter distribution,
which is generally described as

\begin{equation}
\ \rho\,(r) \propto \frac{1}{(\frac{r}{a})^{\gamma} [1 + (\frac{r}{a}%
)^{\alpha} ] ^{(\beta - \gamma)/\alpha}}
\end{equation}
\

Particular values of the parameters give models with a singular
behaviour towards the galactic center, which could determine a
strong enhance of gamma flux in that direction.\newline Other
models postulate a clumped distribution of dark matter, with the
density profile describing the average distribution of dark matter
in the galactic halo. One can expect local enhancement of gamma
radiation from those regions at higher neutrino densities.
\newline The simplest profile is described by an isothermal sphere
for heavy neutrino clustering in a galactic halo with length scale
$a \geq 10\,kpc$ and mass density

\[
\rho = \frac{\rho_0}{1 + (r/a)^2}
\]

where $\rho_0$ is the central density of the halo.\newline
In the spherical model the square density integral leads to an adimensional
intensity $I(\psi)$

\[
I(\psi) = \int \frac{1}{(1 + (r/a)^2)^2} dr(\psi)/a ;
\]

this intensity has a characteristic behaviour which is maximum in the
direction of the galactic center ($\psi = 0$), and then decreases for $0 <
\psi < \pi$, but it doesn't vary more than a factor ten with the angular
coordinate.\newline
At high latitudes $I\,(\psi)$ is generally of order unity.\newline

A rough way to calculate the photon flux by neutrino annihilation
through the channel $N\bar{N} \rightarrow q\bar{q}$ comes from the
analysis of $Z$ decay. The probability of a hadronic Z decay is 20
times greater than its decay in an electron pair ($\Gamma (Z
\rightarrow hadrons)/\Gamma (Z \rightarrow e^-e^+) = 20.795 \pm
0.040$ [27]), with an average production of
$<n_{\pi^0}> \sim 9$ neutral pions. So the cross section of a hadronic $N%
\bar{N}$ annihilation would roughly be $(\sigma v)_{hadr} \sim 20
(\sigma v)_{e^-e^+}$, where $(\sigma v)_{e^-e^+}$ has been
previously calculated [14]. Assuming that the fundamental
contribution to the flux is given by a spherical region of radius
$a$, Eq. (13) becomes

\begin{equation}
\ J_{\gamma} = 2 < n_{\pi^0} > \frac{(\sigma v)_{hadr}}{4\pi} n^2_{0 N}a
I(\psi)
\end{equation}
\

where $n_{0 N} = \rho_0 /m_N$ is neutrino central number
density.\newline With a neutrino mass $m_N = 50\,Gev$ a flux
estimate gives

\begin{equation}
\ J_{\gamma} \sim 1.2\cdot 10^{-6} I(\psi) \left( \frac{\Omega_{clust}}{10^7}
\right) \left( \frac{a}{30\,kpc} \right) \,cm^{-2}\,s^{-1}\,sr^{-1}
\end{equation}
\

where $\Omega_{clust} = (\rho_{G}/\rho_b)^{3/4}$ ($\rho_{G}$ is
the average matter density in the Galaxy, while $\rho_b$ is the
cosmological baryonic matter density) describes the increase in
neutrino density during the clustering in the galaxy. This result
is comparable to EGRET measures (see previous section).\newline
Monte Carlo simulations of neutrino annihilations [19] have been
also used to compare EGRET flux, showing that it is possible to
extrapolate a power law for the gamma spectrum.\newline An
approximated integral flux for $m_N = 50\,GeV$ and $a \sim
10\,kpc$  is roughly

\begin{equation}
\ \Phi_{\gamma} > 6\cdot 10^{-7}\,I(\psi) \,cm^{-2} \,s^{-1} \,sr^{-1},
\end{equation}
\

while for $m_N = 100\,GeV$ at $a \sim 10\,kpc$

\begin{equation}
\ \Phi_{\gamma} > 4\cdot 10^{-7}\,I(\psi) \,cm^{-2} \,s^{-1} \,sr^{-1},
\end{equation}
\

Finally we notice that due to the drastic increase of $\sigma_{N\bar{N}}$ at
Z pole and the consequent suppression of $N\bar{N}$ relic number, the
expected gamma flux at $m_N \simeq 45\,GeV$ is much smaller and negligible,
with

\begin{equation}
\ \Phi_{\gamma} > 2\cdot 10^{-8}\,I(\psi) \,cm^{-2} \,s^{-1} \,sr^{-1},
\end{equation}
\

\section{Conclusions}

The role of a heavy neutrino as a non dominant cold dark matter component
has been analyzed in order to explain the Gamma halo around our Galaxy.%
\newline
We considered a narrow range of neutrino masses between the values
$M_Z/2 < m_N < M_Z$. The lower part of this range has been fixed
as a consequence of last results of DAMA experiment, that show a
consistency of NaI detector signal with a heavy neutrino CDM
candidate, having a mass $45\,GeV < m_N \leq 50 \,GeV$ (with the
upper limit only giving a relic neutrino density high enough to be
cosmologically relevant).\newline Higher mass values are excluded
by the analysis of experimental data up to 300 GeV [6,14].\newline
Two different emission mechanism have been analyzed and compared:
ICS of relativistic electrons (originated in $N\bar{N}$
annihilations) on optic
background photons, and direct neutrino annihilation in high energy photons.%
\newline
Both processes determine a radiation flux close to experimental
results, but distinguish themselves for a different galactic
distribution of the radiation produced.
\begin{enumerate}
  \item  A spherical symmetry in the case of prompt gamma photons by
$N\bar{N}$ annihilations  close to neutrino profile (assuming the
smooth model) in the halo.

  \item  A spherical (neutrino) with a
spheroidal (photon) distribution recalling more the luminous
structure of the visible part of the Galaxy if gamma rays are
generated by ICS.

\end{enumerate}

 The profile of photon flux showed by
Dixon et al. indicates a different morphology of gamma halo at
different energy ranges, with the evidence of a
halo excess not correlated with a component on the galactic plane for $%
300\,MeV < E < 1 \,GeV$, while for $E > 1 \,GeV$ a planar gamma
component is clearly detected.\newline Actually it is not possible
to determine which is the real process source of this high energy
photons (ICS or pion decay), but neutrino annihilations could
describe more easily than other models the presence of a high
galactic latitude emission recently discovered.\newline

In conclusion the possibility to solve at once the DAMA signals
and the gamma halo at GeV with a unique fourth neutrino generation
is fascinating: near forthcoming Lep II data analysis of $e^+ e^-
\rightarrow N \bar{N} \gamma$ events [28] may confirm or exclude
this new physics window.

\section*{Acknowledgements}

We are grateful to Yu. A. Golubkov for interesting discussions and numerical
simulations. M. Khlopov and R. Konoplich are grateful to Rome I University
and Rome III University for hospitality and support.\newline
The work was partly performed in the framework of International Astrodamus
Project with support of Eurocos - AMS and Cosmion - ETHZ collaboration.

%Bibliography

\end{document}